\documentstyle[aps,prd]{revtex}
\tighten

\begin{document}

\draft


\preprint{hep-th/0004067}
\title{Gauge-invariant gravitational perturbations of maximally
symmetric spacetimes} 
\author{Shinji Mukohyama}
\address{
Department of Physics and Astronomy, University of Victoria\\ 
Victoria, BC, Canada V8W 3P6
}

\date{\today}

\maketitle


\begin{abstract} 

Gravitational perturbations of anti-deSitter spacetime play important
roles in AdS/CFT correspondence and the brane world scenario. In this
paper, we develop a gauge-invariant formalism of 
gravitational perturbations of maximally symmetric spacetimes
including anti-deSitter spacetime. Existence of scalar-type master
variables is shown and the corresponding master equations are
derived. 

\end{abstract}



\section{Introduction}
	\label{sec:intro}


Recently, anti-deSitter (AdS) spacetime has been attracting a great
deal of physical interests. In AdS/CFT correspondence, gravitational
theory in AdS background is dual to a conformal field theory (CFT) on
the boundary of the AdS~\cite{Maldacena,AdS-CFT}. It is believed that
a correlation function in the CFT can be calculated by a path integral
in the gravitational theory in AdS background with a certain boundary
condition at the boundary. Moreover, the large $N$ limit of the CFT is
corresponding to the classical limit of the gravitational theory in
AdS, where $N$ is the number of colors. Therefore, the classical 
scattering of gravitational fields in AdS background is an important
issue. In other words, it is important to investigate classical
perturbations of AdS spacetime.


Another subject in which AdS spacetime plays important roles is the
brane-world scenario. Randall and Sundrum~\cite{RS} showed that, in a
$5$-dimensional AdS background, $4$-dimensional Newton's law of 
gravity can be reproduced on a $4$-dimensional timelike hypersurface
despite the existence of the infinite fifth dimension. To be precise,
they considered a $4$-dimensional timelike thin-shell with its tension 
fine-tuned and showed that zero modes of gravitational perturbations
are confined along the thin-shell and are decoupled from all non-zero
modes in low energy. Therefore, it may be possible to consider the
thin-shell, or the world volume of a $3$-brane, as our universe,
provided that matter fields can be confined on the $3$-brane. 
In this respect, many authors investigated validity of the brane-world 
scenario from various points of view. For example, $4$-dimensional
effective Einstein equation on the thin-shell was derived~\cite{SMS};
instability of the Cauchy horizon was discussed by non-linear
analysis~\cite{CG}; gravitational force between two test bodies was 
calculated~\cite{GT,Tanaka-Montes,SSM,GKR}; black holes in the
brane-world were discussed~\cite{CHR,EHM}; inflating brane solutions
were constructed~\cite{Nihei,Kaloper,KK}. Relations to the AdS/CFT
correspondence were also discussed~\cite{Gubser}. In all of these
works, AdS spacetime or its modifications play important roles.


Moreover, recently, cosmological solutions in this scenario were
found~\cite{CGS,FTW,BDEL,Mukohyama,Vollic,Kraus,Ida}. In these 
solutions, the standard cosmology is restored in low energy, provided 
that a parameter $\mu$ in the solutions is small enough. If the
parameter $\mu$ is not small enough, it affects cosmological evolution 
of our universe as dark radiation~\cite{Mukohyama}. Hence, the
parameter $\mu$ should be very small in order that the brane-world 
scenario should be consistent with nucleosynthesis~\cite{BDEL}. On the 
other hand, in Ref.~\cite{MSM}, it was shown that $5$-dimensional
geometry of all these cosmological solutions is the Schwarzschild-AdS
(Sch-AdS) spacetime~\cite{Birmingham} and that $\mu$ is the mass
parameter of the black hole. Therefore, the $5$-dimensional bulk
geometry should be the Sch-AdS spacetime with a small mass, which is
close to the pure AdS spacetime. Moreover, black holes with small mass
will evaporate in a short timescale~\cite{Hawking}. Thus, it seems a
good approximation to consider the pure AdS spacetime as a
$5$-dimensional bulk geometry for the brane-world cosmology.


Since the cosmological solution reproduces the standard cosmology as 
evolution of a homogeneous isotropic universe in low energy, this
scenario may be considered as a realistic cosmology. Hence, it seems
effective to look for observable consequences of this scenario. For
this purpose, cosmic microwave background (CMB) anisotropy is a
powerful tool. Therefore, we would like to give theoretical
predictions of the brane-world scenario on the CMB
anisotropy. However, this is not a trivial task as we shall explain
below~\footnote{A part of information about the CMB anisotropy can be
derived from the conservation of energy momentum
tensor~\cite{conservation}.}.
The main points are the following two: (i) how to give the initial
condition; (ii) how to evolve perturbations. As for the first point,
there is essentially the same issue even in the standard
cosmology. The creation-from-nothing scenario may solve 
it~\cite{GS,Koyama-Soda,HHR} or may not.


Regarding the second point, we would like to argue that evolution of
cosmological perturbations becomes non-local in the brane-world
scenario. First, provided a suitable initial condition is given,
perturbations localized on the brane can produce gravitational waves. 
Next, those gravitational waves propagate in the bulk AdS spacetime,
and may collide with the brane at a spacetime point different from the 
spacetime point at which the gravitational waves were produced. 
When they collide with the brane, they should alter evolution of
perturbations localized on the brane. Hence, evolution of
perturbations localized on the brane should be non-local in the sense
that it should be described by some integro-differential
equations. Thus, the non-locality seems the essential point of
evolution of cosmological perturbations in the brane world
scenario. Without considering this point, we cannot expect drastic
differences between CMB anisotropies predicted by the brane-world
cosmology and the standard cosmology. Therefore, we have to consider
the non-locality caused by gravitational waves seriously.


Towards the derivation of the integro-differential equations, it seems
an important step to analyze gravitational perturbations in the bulk
AdS geometry.


The purpose of this paper is to develop a gauge-invariant formalism of
gravitational perturbations of maximally symmetric spacetimes
including AdS spacetime. Existence of scalar-type master variables is
shown, and the corresponding master equations are derived.


In Sec.~\ref{sec:background} properties of the background spacetime
are summarized. In Sec.~\ref{sec:variables} gauge-invariant variables
are constructed. In Sec.~\ref{sec:Einstein-eq} linearized Einstein
equation is expressed in terms of the gauge-invariant variables. In
Sec.~\ref{sec:master-eq} existence of scalar-type master variables is
shown and the corresponding master equations are derived. 
Sec.~\ref{sec:summary} is devoted to a summary of this paper.


\section{Background spacetime}
	\label{sec:background}


A spacetime is said to be maximally symmetric if it admits the maximum 
number $D(D+1)/2$ of independent Killing vector fields, where $D$ is
the dimensionality of the spacetime. It can be shown that a maximally
symmetric spacetime is a spacetime of constant curvature and that it
is uniquely specified by a curvature constant~\cite{Weinberg}. These
are deSitter, Minkowski, and anti-deSitter spacetimes for positive,
zero, and negative values of the curvature constant,
respectively~\cite{Hawking-Ellis}.

Since we consider a maximally symmetric spacetime as the background
geometry and it is of constant curvature as mentioned above, we have
the following equation for the background curvature tensor. 
%
\begin{equation}
 R^{(0)}_{MLNL'} = 
	\frac{2\Lambda}{(D-1)(D-2)}
	(g^{(0)}_{MN}g^{(0)}_{LL'} - g^{(0)}_{ML'}g^{(0)}_{LN}), 
	\label{eqn:const-curv}
\end{equation}
where $\Lambda$ is a constant called a cosmological constant, and the
superscript $(0)$ hereafter denotes that the quantity is calculated
for the unperturbed spacetime. Note that the normalization in the
right hand side is determined so that 
%
\begin{equation}
 G^{(0)}_{MN} + \Lambda g^{(0)}_{MN} = 0. 
\end{equation}


As the brane-world cosmology, in many cases of physical interests, the
boundary of an unperturbed spacetime is a world-volume of a
constant-curvature (or equivalently, maximally symmetric)
subspace. Hence, it is convenient to decompose the background
spacetime into a family of constant-curvature subspace: let us
consider the decomposition  
%
\begin{equation}
 g^{(0)}_{MN} = \gamma_{ab}dx^adx^b+r^2\Omega_{ij}dx^idx^j,
	\label{eqn:gamma-Omega}
\end{equation}
where $\Omega_{ij}$ is a metric of a $n$-dimensional
constant-curvature space, $\gamma_{ab}$ is a $(D-n)$-dimensional
metric depending only on the $(D-n)$-dimensional coordinates
$\{x^a\}$, and $r$ also depends only on $\{x^a\}$. Denoting the
curvature constant of $\Omega_{ij}$ by $K$, we can write the curvature 
tensor of $\Omega_{ij}$ as
%
\begin{equation}
 R^{(\Omega)ij}_{\qquad kl} = 
	K(\delta^i_k\delta^j_l - \delta^i_l\delta^j_k). 
	\label{eqn:R-Omega}
\end{equation}
By using this expression, it is easy to show by explicit calculation
that the curvature tensor of the background metric of the form
(\ref{eqn:gamma-Omega}) has the following components. 
%
\begin{eqnarray}
 R^{(0)ij}_{\qquad kl} & = & \left(
	\frac{K}{r^2}-\gamma^{ab}\partial_a\ln r \partial_b\ln r
	\right)(\delta^i_k \delta^j_l - \delta^i_l \delta^j_k),
	\nonumber\\
 R^{(0)i}_{\quad\ ajb} & = & -\delta^i_j
	(\nabla_a\nabla_b\ln r + \partial_a\ln r\partial_b\ln r), 
	\nonumber\\
 R^{(0)}_{abcd} & = & R^{(\gamma)}_{abcd}, 
\end{eqnarray}
where $\nabla_a$ is the covariant derivative compatible with the
metric $\gamma_{ab}$, and $R^{(\gamma)}_{abcd}$ is the curvature
tensor of $\gamma_{ab}$. Therefore, the condition
(\ref{eqn:const-curv}) implies that 
%
\begin{eqnarray}
 \gamma^{ab}\partial_a\ln r \partial_b\ln r & = & 
	\frac{K}{r^2} - \frac{2\Lambda}{(D-1)(D-2)},
	\nonumber\\
 \nabla_a\nabla_b\ln r + \partial_a\ln r\partial_b\ln r & = & 
	- \frac{2\Lambda\gamma_{ab}}{(D-1)(D-2)},
	\nonumber\\
 R^{(\gamma)}_{abcd} & = & \frac{2\Lambda}{(D-1)(D-2)}
	(\gamma_{ac}\gamma_{bd} - \gamma_{ad}\gamma_{bc}).
	\label{eqn:const-curv-dec}
\end{eqnarray}
These equations will be used repeatedly in this paper.


Now, we have three kinds of covariant derivatives for the background
geometry: the first, which we shall denote by semicolons, is the
covariant derivative compatible with the original $D$-dimensional
background metric $g^{(0)}_{MN}$; secondly, $\nabla_a$ is the
covariant derivative compatible with the $(D-n)$-dimensional metric
$\gamma_{ab}$; and the third, which we shall denote by $D_{i}$, is
compatible with $\Omega_{ij}$. The following examples of relations
among them can be easily obtained. First, for an arbitrary 1-form
field $V_M$, 
%
\begin{eqnarray}
 V_{a;b} & = & \nabla_b V_a, \nonumber\\
 V_{a;i} & = & \partial_iV_a - V_i\partial_a\ln r,\nonumber\\
 V_{i;a} & = & \partial_aV_i - V_i\partial_a\ln r,\nonumber\\
 V_{i;j} & = & D_jV_i + r^2\Omega_{ij}\gamma^{ab}V_a\partial_b\ln r. 
	\label{eqn:derivatives}
\end{eqnarray}
These relations will be used when we seek infinitesimal gauge
transformations for perturbations in Sec.~\ref{sec:variables}. The
second set of examples is given in Appendix~\ref{app:derivatives} and
will be useful when we calculate Einstein equation for perturbations
in Sec.~\ref{sec:Einstein-eq}.


\section{Gauge-invariant variables}
	\label{sec:variables}

The main purpose of this paper is to derive equations governing
gravitational perturbations around the background specified in the
previous section. In other words, defining perturbation 
$\delta g_{MN}$ by 
%
\begin{equation}
 g_{MN} = g^{(0)}_{MN} + \delta g_{MN}, 
\end{equation}
we shall derive the Einstein equation linearized with respect to
$\delta g_{MN}$. Hence, in principle, independent variables are all
components of $\delta g_{MN}$. However, those include degrees of 
freedom of gauge transformation as we shall see explicitly. Thus, it
is desirable to reduce the number of degrees of freedom so that
reduced degrees of freedom include physical perturbations only. For
this purpose, in this section, we shall construct gauge invariant
variables. The advantages of this approach against gauge-fixing will
be explained in Sec.~\ref{sec:summary}.


Since the background geometry still has the symmetry of isometry of
$\Omega_{ij}$ even after the decomposition (\ref{eqn:gamma-Omega}), it 
is convenient to expand perturbations by harmonics on the
constant-curvature space:
%
\begin{eqnarray}
 \delta g_{MN} dx^M dx^N & = & \sum_k\left[ h_{ab}Ydx^adx^b 
	+ 2(h_{(T)a}V_{(T)i}+h_{(L)a}V_{(L)i})dx^adx^i\right.
	\nonumber\\
 & & 	\left.+ (h_{(T)}T_{(T)ij}+h_{(LT)}T_{(LT)ij}
	+h_{(LL)}T_{(LL)ij}+h_{(Y)}T_{(Y)ij})dx^idx^j\right],
	\label{eqn:expansion-deltag}
\end{eqnarray}
where $Y$, $V_{(T,L)}$ and $T_{(T,LT,LL,Y)}$ are scalar, vector and
tensor harmonics, respectively, and the coefficients $h_{ab}$,
$h_{(T,L)a}$ and $h_{(T,LT,LL,Y)}$ are supposed to depend only on the
$(D-n)$-dimensional coordinates $\{x^a\}$. 
Hereafter, $k$ denotes continuous ($K=0,-1$) or discrete ($K=1$)
eigenvalues, and we omit them in most cases. In this respect, the 
summation with respect to $k$ should be understood as an integration
for $K=0,-1$. See Appendix~\ref{app:harmonics} for definitions and
basic properties of the harmonics.


As mentioned already, $\delta g_{MN}$ includes degrees of freedom of
gauge transformation. In fact, an infinitesimal gauge transformation
is given by
%
\begin{equation}
 \delta g_{MN} \to \delta g_{MN} 
	- \bar{\xi}_{M;N} - \bar{\xi}_{N;M},
\end{equation}
where $\bar{\xi}_M$ is an arbitrary vector field. Hence, by expanding
the vector $\bar{\xi}_M$ in terms of harmonics as
%
\begin{equation}
 \bar{\xi}_Mdx^M = \sum_k\left[\xi_aYdx^a + 
	(\xi_{(T)}V_{(T)i}+\xi_{(L)}V_{(L)i})dx^i\right],
\end{equation}
we get the following infinitesimal gauge transformation for the
expansion coefficients in Eq.~(\ref{eqn:expansion-deltag}). 
%
\begin{eqnarray}
 h_{ab} & \to & 
	h_{ab} - \nabla_a\xi_b - \nabla_b\xi_a,
	\nonumber\\
 h_{(T)a} & \to & 
	h_{(T)a} - r^2\partial_a(r^{-2}\xi_{(T)}),
	\nonumber\\
 h_{(L)a} & \to & 
	h_{(L)a} - \xi_a - r^2\partial_a(r^{-2}\xi_{(L)}),
	\nonumber\\
 h_{(T)} & \to & 
	h_{(T)},
	\nonumber\\
 h_{(LT)} & \to & 
	h_{(LT)} - \xi_{(T)},
	\nonumber\\
 h_{(LL)} & \to & 
	h_{(LL)} - \xi_{(L)},
	\nonumber\\
 h_{(Y)} & \to & 
	h_{(Y)} - \gamma^{ab}\xi_a\partial_br^2 
	+ \frac{2k^2}{n}\xi_{(L)}. 
	\label{eqn:gauge-tr}
\end{eqnarray}
Note that we have used Eq.~(\ref{eqn:derivatives}) to derive those 
gauge transformations.


From the gauge transformations~(\ref{eqn:gauge-tr}), it is easy to
construct gauge invariant variables as follows.
%
\begin{eqnarray}
 F_{ab} & = & h_{ab} - \nabla_aX_b - \nabla_bX_a, 
	\nonumber\\
 F & = & h_{(Y)} - \gamma^{ab}X_a\partial_br^2 
	+ \frac{2k^2}{n}h_{(LL)}, 
	\nonumber\\
 F_a & = & h_{(T)a} - r^2\partial_a(r^{-2}h_{(LT)}),
	\nonumber\\
 F_{(T)} & = & h_{(T)},
\end{eqnarray}
where $X_a$ is a gauge-dependent combination defined by
%
\begin{equation}
 X_a = h_{(L)a} - r^2\partial_a(r^{-2}h_{(LL)}),
\end{equation}
and transforms under the infinitesimal gauge transformation as
%
\begin{equation}
 X_a \to X_a - \xi_a. 
\end{equation}

Note that, for $k^2=0$, $F_{ab}$ and $F$ are not gauge invariant
variables but gauge dependent variables since 
$V_{(L)i}\equiv 0$ and $T_{(LL)ij}\equiv 0$. Similarly, for a special
value of $k$ such that $T_{(LT)ij}\equiv 0$, $F_{a}$ is not a gauge
invariant variable but a gauge dependent variable. For all other
values of $k$, off course, $F_{ab}$, $F$, $F_{a}$ and $F_{(T)}$ are
gauge invariant variables.


\section{Einstein equation for the gauge-invariant variables}
	\label{sec:Einstein-eq}

In this section we shall seek linearized equations for the gauge
invariant variables $F_{ab}$, $F$ and $F_{(T)}$ constructed in the
previous section. For this purpose, first, we expand the Einstein
tensor in powers of $\delta g_{MN}$ without using the expansion
(\ref{eqn:expansion-deltag}) nor any properties of the background
geometry. The result is 
%
\begin{equation}
 G_{MN} = G^{(0)}_{MN} + G^{(1)}_{MN} + O(\delta g^2),
\end{equation}
where
%
\begin{eqnarray}
 2 G^{(1)}_{MN} & = & -\delta g_{MN;L}^{\quad\ ;L} 
	+ ( \delta g_{\ M;LN}^L + \delta g_{\ N;LM}^L )
	- \delta g_{;MN}
	+ ( \delta g^{;L}_{\ ;L} - \delta g^{LL'}_{\quad ;LL'} )
	g^{(0)}_{MN} \nonumber\\
 & & 	+ ( R^{(0)L}_M\delta g_{LN} + R^{(0)L}_N\delta g_{LM} )
	+ g^{(0)}_{MN} R^{(0)}_{LL'}\delta g^{LL'}
	- R^{(0)} \delta g_{MN} - 2 R^{(0)}_{MLNL'} \delta g^{LL'},
\end{eqnarray}
and $\delta g^{MN}\equiv g^{(0)MM'}g^{(0)NN'}\delta g_{M'N'}$, 
$\delta g\equiv g^{(0)MN}\delta g_{MN}$. 
Correspondingly, the linearized Einstein equation becomes
%
\begin{equation}
  G^{(1)}_{MN} + \Lambda \delta g_{MN} = 0. 
	\label{eqn:lin-Ein-eq}
\end{equation}
Next, because of the constant-curvature condition
(\ref{eqn:const-curv}) for the background, the left hand side of
Eq.~(\ref{eqn:lin-Ein-eq}) multiplied by $2$ is rewritten as
%
\begin{eqnarray}
 2 ( G^{(1)}_{MN} + \Lambda \delta g_{MN} ) & = & 
	-\delta g_{MN;L}^{\quad\ ;L} 
	+ ( \delta g_{\ M;LN}^L + \delta g_{\ N;LM}^L )
	- \delta g_{;MN}
	+ ( \delta g^{;L}_{\ ;L} - \delta g^{LL'}_{\quad ;LL'} )
	g^{(0)}_{MN} \nonumber\\
 & & 	+ \frac{4\Lambda}{(D-1)(D-2)}\delta g_{MN}
	+ \frac{2(D-3)\Lambda}{(D-1)(D-2)}\delta g\ g^{(0)}_{MN}. 
	\label{eqn:G1+Ldg}
\end{eqnarray}
Thirdly, by using the formulas (\ref{eqn:2ndderivatives}) given in
Appendix~\ref{app:derivatives} and substituting the expansion
(\ref{eqn:expansion-deltag}) into Eq.~(\ref{eqn:G1+Ldg}), we obtain
the following expansion of the linearized Einstein equation. 
%
\begin{eqnarray}
 2(G^{(1)}_{MN}+\Lambda \delta g_{MN})dx^Mdx^N & = & 
	\sum_k\left[ E_{ab}Ydx^adx^b 
	+ 2(E_{(T)a}V_{(T)i}+E_{(L)a}V_{(L)i})dx^adx^i\right.
	\nonumber\\
 & & 	\left.+ (E_{(T)}T_{(T)ij}+E_{(LT)}T_{(LT)ij}
	+E_{(LL)}T_{(LL)ij}+E_{(Y)}T_{(Y)ij})dx^idx^j\right],
\end{eqnarray}
where the coefficients $E_{ab}$, $E_{(T,L)a}$ and $E_{(T,LT,LL,Y)}$
depend only on the $(D-n)$-dimensional coordinates
$\{x^a\}$. Hereafter, let us call those coefficients 
{\it Einstein-equation-forms}. The linearized Einstein equation is
equivalent to the following set of projected equations. 
%
\begin{eqnarray}
 E_{ab} & = & 0,	\nonumber\\
 E_{(T)a} & = & E_{(L)a} = 0,	\nonumber\\
 E_{(T)} & = & E_{(LT)} = E_{(LL)} = E_{(Y)} = 0. 
\end{eqnarray}
The next task is to express all Einstein-equation-forms in terms of
the gauge invariant variables only. However, before showing the
results, let us make classification of perturbations in order to
make arguments clear.

Now, even without explicit expressions, it is easily shown from
orthogonality between different kinds of harmonics (see
Appendix~\ref{app:harmonics}) that 
(i) $E_{ab}$, $E_{(L)a}$, $E_{(LL)}$ and $E_{(Y)}$ depend only on 
$h_{ab}$, $h_{(L)a}$, $h_{(LL)}$ and $h_{(Y)}$; 
(ii) $E_{(T)a}$ and $E_{(LT)}$ depend only on $h_{(T)a}$ and
$h_{(LT)}$; 
(iii) $E_{(T)}$ depends only on $h_{(T)}$. 
Therefore, it is convenient to classify all perturbations into three
categories: 
(i) ($h_{ab}$, $h_{(L)a}$, $h_{(LL)}$, $h_{(Y)}$); 
(ii) ($h_{(T)a}$, $h_{(LT)}$);
(iii) $h_{(T)}$. 
It is evident that each category can be analyzed independently. Let us 
call perturbations in the first, second and third categories {\it
scalar perturbations}, {\it vector perturbations} and 
{\it tensor perturbations}, respectively~\footnote{
This way of classification is the same as that adopted in the theory
of cosmological perturbations~\cite{Kodama-Sasaki}}.

\subsection{Einstein equation for scalar perturbations}

For scalar perturbations given by 
%
\begin{equation}
 \delta g_{MN} dx^M dx^N = \sum_k\left[ h_{ab}Ydx^adx^b 
	+ 2h_{(L)a}V_{(L)i}dx^adx^i 
	+ (h_{(LL)}T_{(LL)ij}+h_{(Y)}T_{(Y)ij})dx^idx^j\right],
\end{equation}
appropriate gauge invariant variables are $F_{ab}$ and $F$, and
appropriate Einstein-equation-forms are $E_{ab}$, $E_{(L)a}$,
$E_{(LL)}$ and $E_{(Y)}$. The explicit expressions for the
Einstein-equation-forms in terms of the gauge invariant variables are
as follows. 
%
\begin{eqnarray}
 E_{ab} & = & -\nabla^2F_{ab}+\nabla_a\nabla^cF_{cb}
	+\nabla_b\nabla^cF_{ca}-\nabla_a\nabla_bF^c_c	\nonumber\\
 & & 	+n(\nabla_aF_b^c+\nabla_bF_a^c-\nabla^cF_{ab})\partial_c\ln r
	+\left[\frac{k^2}{r^2}-\frac{4(n-1)\Lambda}{(D-1)(D-2)}\right]
	F_{ab}	\nonumber\\
 & & 	+ \frac{n}{r^2}\left\{
	-\nabla_a\nabla_bF + \partial_aF\partial_b\ln r
	+\partial_bF\partial_a\ln r - 2F\partial_a\ln r\partial_b\ln r
	\right\}	\nonumber\\
 & & 	+\gamma_{ab}\left\{\nabla^2F^c_c-\nabla^c\nabla^dF_{cd}
	+n(\nabla^dF^c_c-2\nabla_cF^{cd})\partial_d\ln r 
	\right.\nonumber\\
 & & 	\left.+\left[\frac{2(D-3)\Lambda}{(D-1)(D-2)}-\frac{k^2}{r^2}
	\right]F^c_c -nF^{cd}(n\partial_c\ln r\partial_d\ln r
	+\nabla_c\nabla_d\ln r)\right\}	\nonumber\\
 & & 	+\frac{\gamma_{ab}}{r^2}\left\{n\nabla^2F
	+n(n-3)\partial^cF\partial_c\ln r
	\right. 	\nonumber\\
 & & 	\left.+F\left[ -n(n-2)\partial^c\ln r\partial_c\ln r 
	- n\nabla^2\ln r + \frac{2(D-3)n\Lambda}{(D-1)(D-2)}
	- (n-1)\frac{k^2}{r^2}\right]\right\},
	\nonumber\\
 E_{(L)a} & = & r^{-(n-2)}\nabla^b(r^{n-2}F_{ab})
	- (n-1)\partial_a(r^{-2}F) - r\partial_a(r^{-1}F^b_b)
	,\nonumber\\
 E_{(LL)} & = & -\frac{1}{2}[F^a_a+(n-2)r^{-2}F],\nonumber\\
 E_{(Y)} & = & nr^2\left\{\nabla^2F^a_a-\nabla^a\nabla^bF_{ab}
	-2(n-1)\nabla_aF^{ab}\partial_b\ln r
	+(n-1)\partial^bF^a_a\partial_b\ln r \right.\nonumber\\
 & & 	\left.-F^{ab}\left[(n^2-2n+2)\partial_a\ln r\partial_b\ln r+ 
	n\nabla_a\nabla_b\ln r\right]
	+F^a_a\left[\frac{2(D-3)\Lambda}{(D-1)(D-2)} 
	-\frac{n-1}{n}\frac{k^2}{r^2}\right]\right\}\nonumber\\
 & & 	+ n\left\{(n-1)\nabla^2F+(n-4)(n-1)\partial^aF\partial_a\ln r
	\right.\nonumber\\
 & & 	\left.+ F\left[-(n^2-4n+2)\partial^a\ln r\partial_a\ln r
	-(n-2)\nabla^2\ln r + \frac{2(Dn-3n+2)\Lambda}{(D-1)(D-2)}
	-\frac{(n-1)(n-2)}{n}\frac{k^2}{r^2}\right]\right\}.
	\label{eqn:Ein-eq-form-scalar}
\end{eqnarray}
We have used the relations (\ref{eqn:const-curv-dec}) to derive these
expressions. Note that these are expressed in terms of gauge-invariant 
variables only, as expected.

Although each of the Einstein-equation-forms gives an equation for
scalar perturbations, they do not give independent equations because
of the Bianchi identity. In fact, the equation $E_{(Y)}=0$ can be
derived from $E_{(L)a}=0$ and $E_{(LL)}=0$. Thus, a set of independent
equations of motion for scalar perturbations are given by $E_{ab}=0$
and 
%
\begin{eqnarray}
 F^a_a + (n-2)r^{-2}F & = & 0,\nonumber\\
 \nabla^b(r^{n-2}F_{ab}) & = & \partial_a(r^{n-4}F).
	\label{eqn:Ein-scalar}
\end{eqnarray}
Here we have rewritten $E_{(L)a}=0$ into the form of the last equation 
by using $E_{(LL)}=0$.

\subsection{Einstein equation for vector perturbations}

For vector perturbations given by 
%
\begin{equation}
 \delta g_{MN}dx^Mdx^N = \sum_k\left[
	2h_{(T)a}V_{(T)i}dx^adx^i 
	+ h_{(LT)}T_{(LT)ij}dx^idx^j\right],
\end{equation}
the appropriate gauge invariant variable is $F_{a}$, and appropriate
Einstein-equation-forms are $E_{(T)a}$ and $E_{(LT)}$. The explicit 
expressions for the Einstein-equation-forms in terms of the gauge
invariant variables are as follows. 
%
\begin{eqnarray}
 E_{(T)a} & = & -r^{-n}\nabla^b\left[ 
	r^{n+2}\nabla_b\left(\frac{F_a}{r^2}\right)
	- r^{n+2}\nabla_a\left(\frac{F_b}{r^2}\right)\right]
	+ \frac{k^2-(n-1)K}{r^2}F_a, \nonumber\\
 E_{(LT)} & = & r^{-(n-2)}\nabla^a(r^{n-2}F_a). 
	\label{eqn:Ein-eq-form-vec}
\end{eqnarray}
We have used the relations (\ref{eqn:const-curv-dec}) to derive these
expressions.

\subsection{Einstein equation for tensor perturbations}

For tensor perturbations given by 
%
\begin{equation}
 \delta g_{MN}dx^Mdx^N = \sum_k h_{(T)}T_{(T)ij}dx^idx^j, 
\end{equation}
the coefficient $h_{(T)}$ itself is the gauge invariant variable
$F_{(T)}$, and the appropriate Einstein-equation-form is
$E_{(T)}$. The explicit expressions for the Einstein-equation-form is
given by 
%
\begin{equation}
 E_{(T)} = -r^{-(n-2)}\nabla^a[r^n\nabla_a(r^{-2}F_{(T)})]
	+ \frac{k^2+2K}{r^2}F_{(T)}, 
\end{equation}
or equivalently, 
%
\begin{equation}
 E_{(T)} = -r^{D-n+1}\nabla^a[r^{-(2D-n-2)}\nabla_a(r^{D-3}F_{(T)})] 
	+ \frac{k^2+[(D-1)(n-2)-D(D-3)]K}{r^2}F_{(T)}. 
\end{equation}
We have used the relations (\ref{eqn:const-curv-dec}) to derive these
expressions.


\section{Master equations}
	\label{sec:master-eq}

In the previous section we obtained equations of motion for the gauge
invariant variables. These are described as scalars ($F$ and
$F_{(T)}$), vectors ($F_a$) and $2$-tensors ($F_{ab}$) on the
$(D-n)$-dimensional spacetime with the metric $\gamma_{ab}$. In the
easiest case when $D-n=1$, those vectors and tensors have only one
component, and thus they can trivially be treated on the same footing
as scalars. However, in general, treatment of vectors and tensors is 
more complicated than scalars. In this section we show that, also in
the case when $D-n=2$, those vector fields and tensor fields can be
described by scalar fields called master variables.

Since we consider only the $D-n=2$ case in this section, without loss
of generality, we can adopt the following form of the metric
$\gamma_{ab}$. 
%
\begin{equation}
 \gamma_{ab}dx^adx^b = -2 e^{\phi}dx_+dx_-,
\end{equation}
where $\phi$ is a function of the coordinates $x_+$ and $x_-$. In this
coordinate, the condition (\ref{eqn:const-curv-dec}) is written as 
%
\begin{eqnarray}
 e^{-\phi}\frac{\partial_+r\partial_-r}{r^2} & = & 
	\frac{\Lambda}{(D-1)(D-2)} - \frac{K}{2r^2}, \nonumber\\
 \partial_+(e^{-\phi}\partial_+r) & = & 0,\nonumber\\
 \partial_-(e^{-\phi}\partial_-r) & = & 0,\nonumber\\
 e^{-\phi}\frac{\partial_+\partial_-r}{r} & = & 
	\frac{2\Lambda}{(D-1)(D-2)}, \nonumber\\
 e^{-\phi}\partial_+\partial_-\phi & = & \frac{2\Lambda}{(D-1)(D-2)}
	\label{eqn:const-curv-dec2}
 \end{eqnarray}

\subsection{Master equation for scalar perturbations}


Now let us show that the tensor $F_{ab}$ and the scalar $F$ can be
described by one scalar variable if they satisfy
Eqs.~(\ref{eqn:Ein-scalar}). First, by defining two scalars
$\Phi_{(S)\pm}$ by 
%
\begin{eqnarray}
 \tilde{F}_{++} & = & 
	e^{\phi}\partial_+(e^{-\phi}\partial_+\Phi_{(S)+}),
	\nonumber\\
 \tilde{F}_{--} & = & 
	e^{\phi}\partial_-(e^{-\phi}\partial_-\Phi_{(S)-}),
\end{eqnarray}
Eqs.~(\ref{eqn:Ein-scalar}) can be rewritten as
%
\begin{equation}
 e^{-\phi}\tilde{F}_{+-} = c \tilde{F},
\end{equation}
and
%
\begin{eqnarray}
 \partial_+[(c+1)\tilde{F}
	+e^{-\phi}\partial_+\partial_-\Phi_{(S)+}
	-\tilde{\Lambda}\Phi_{(S)+}] & = & 0, \nonumber\\
 \partial_-[(c+1)\tilde{F}
	+e^{-\phi}\partial_+\partial_-\Phi_{(S)-}
	-\tilde{\Lambda}\Phi_{(S)-}] & = & 0,
	\label{eqn:dF+dddP-dLP}
\end{eqnarray}
where $\tilde{F}_{ab}=r^{D-4}F_{ab}$, $\tilde{F}=r^{D-6}F$,
$c=(D-4)/2$ and
$\tilde{\Lambda}=2\Lambda/(D-1)(D-2)$. Eqs.(\ref{eqn:dF+dddP-dLP})
imply that there exist functions $f_1(x_-)$ and $f_2(x_+)$ such that 
%
\begin{eqnarray}
 (c+1)\tilde{F}+e^{-\phi}\partial_+\partial_-\Phi_{(S)+}
	-\tilde{\Lambda}\Phi_{(S)+} & = & f_1(x_-), \nonumber\\
 (c+1)\tilde{F}
	+e^{-\phi}\partial_+\partial_-\Phi_{(S)-}
	-\tilde{\Lambda}\Phi_{(S)-} & = & f_2(x_+). 
\end{eqnarray}
Thus, consistency between these two equations requires that 
%
\begin{equation}
 e^{-\phi}\partial_+\partial_-(\Phi_{(S)+}-\Phi_{(S)-}) = 
	\tilde{\Lambda}(\Phi_{(S)+}-\Phi_{(S)-}) + f_1(x_-)-f_2(x_+).
	\label{eqn:consitency}
\end{equation}

Next, let us solve the consistency condition (\ref{eqn:consitency})
explicitly.

When $\tilde{\Lambda}=0$, the last equation of
(\ref{eqn:const-curv-dec2}) implies that there exist functions
$\phi_+$ and $\phi_-$ such that 
$\phi=\phi_+(x_+)+\phi_-(x_-)$. Thus,
(\ref{eqn:consitency}) can be solved easily to give 
%
\begin{equation}
 \Phi_{(S)+}-\Phi_{(S)-} = 
	\bar{x}_+\int dx_- e^{\phi_-(x_-)}f_1(x_-) 
	-\bar{x}_-\int dx_+ e^{\phi_+(x_+)}f_2(x_+) 
	+ f_3(x_-) - f_4(x_+), 
\end{equation}
where $\bar{x}_{\pm}=\int dx_{\pm}e^{\phi_{\pm}(x_{\pm})}$, and 
$f_3$ and $f_4$ are arbitrary functions. 
Therefore, defining $\Phi_{(S)}$ by
%
\begin{eqnarray}
 \Phi_{(S)} & = & \Phi_{(S)+}
	- \tilde{x}_+\int dx_- e^{\phi_-(x_-)}f_1(x_-) - f_3(x_-)
	\nonumber\\
 & = & \Phi_{(S)-} 
	- \tilde{x}_-\int dx_+ e^{\phi_+(x_+)}f_2(x_+) - f_4(x_+),
\end{eqnarray}
$\tilde{F}_{ab}$ and $\tilde{F}$ are written as
%
\begin{eqnarray}
 \tilde{F}_{++} & = & 
	e^{\phi}\partial_+(e^{-\phi}\partial_+\Phi_{(S)}),
	\nonumber\\
 \tilde{F}_{--} & = & 
	e^{\phi}\partial_-(e^{-\phi}\partial_-\Phi_{(S)}),
	\nonumber\\
 e^{-\phi}\tilde{F}_{+-} & = & c\tilde{F} 
	= -\frac{c}{c+1}e^{-\phi}\partial_+\partial_-\Phi_{(S)}.
\end{eqnarray}

On the other hand, when $\tilde{\Lambda}\ne 0$, by defining $\Delta$
by 
$\Delta\equiv
(\Phi_{(S)+}-\Phi_{(S)-})+(f_1(x_-)-f_2(x_+))/\tilde{\Lambda}$, 
the consistency condition (\ref{eqn:consitency}) can be written as 
%
\begin{equation}
 \partial_+\partial_-\Delta = \tilde{\Lambda}e^{\phi}\Delta. 
	\label{eqn:ddD=LeD}
\end{equation}
In Appendix~\ref{app:solution} it is shown that a general solution of
this equation is 
%
\begin{equation}
 \Delta = e^{-\phi}\partial_+(e^{\phi}C^+(x_+))
	+e^{-\phi}\partial_-(e^{\phi}C^-(x_-)), 
\end{equation}
where $C^{\pm}$ are arbitrary functions. Therefore, defining
$\Phi_{(S)}$ by 
%
\begin{eqnarray}
 \Phi_{(S)} & = & \Phi_{(S)+}
	+ f_1(x_-)/\tilde{\Lambda} 
	- e^{-\phi}\partial_-(e^{\phi}C^-(x_-))
	\nonumber\\
 & = & \Phi_{(S)-}
	+ f_2(x_+)/\tilde{\Lambda} 
	+ e^{-\phi}\partial_+(e^{\phi}C^+(x_+)), 
\end{eqnarray}
$\tilde{F}_{ab}$ and $\tilde{F}$ are written as
%
\begin{eqnarray}
 \tilde{F}_{++} & = & 
	e^{\phi}\partial_+(e^{-\phi}\partial_+\Phi_{(S)}),
	\nonumber\\
 \tilde{F}_{--} & = & 
	e^{\phi}\partial_-(e^{-\phi}\partial_-\Phi_{(S)}),
	\nonumber\\
 e^{-\phi}\tilde{F}_{+-} & = & c\tilde{F} 
	= \frac{c}{c+1}(-e^{-\phi}\partial_+\partial_-\Phi_{(S)}
	+ \tilde{\Lambda} \Phi_{(S)}).
	\label{eqn:Fab-F-P}
\end{eqnarray}

In summary, for any value of $\tilde{\Lambda}$ there exists a master
variable $\Phi_{(S)}$ such that $\tilde{F}_{ab}$ and $\tilde{F}$ are
written as (\ref{eqn:Fab-F-P}). These equations can be written
covariantly as 
%
\begin{eqnarray}
 r^{D-4}F_{ab} & = &
	\nabla_a\nabla_b\Phi_{(S)} 
	-\frac{D-3}{D-2}\nabla^2\Phi_{(S)}\gamma_{ab}
	- \frac{2(D-4)\Lambda}{(D-1)(D-2)^2}\Phi_{(S)}\gamma_{ab}
	\nonumber\\
 r^{D-6}F & = & \frac{1}{D-2}\left[\nabla^2\Phi_{(S)}
	+\frac{4\Lambda}{(D-1)(D-2)}\Phi_{(S)}\right].
	\label{eqn:Phi(s)}
\end{eqnarray}


Now, let us derive an equation of motion for the master variable
$\Phi_{(S)}$. First, by substituting expressions (\ref{eqn:Phi(s)})
into the Einstein-equation-form $E_{ab}$ given by
(\ref{eqn:Ein-eq-form-scalar}), we can show that
%
\begin{equation}
 r^{D-2}(E_{ab} - E^c_c\gamma_{ab}) = 
	\nabla_a\nabla_b\Delta_{(S)} + 
	\frac{2\Lambda}{(D-1)(D-2)}\gamma_{ab}\Delta_{(S)}, 
\end{equation}
where
%
\begin{equation}
 \Delta_{(S)} = r^2\left[-\nabla^2\Phi_{(S)} 
	+ (D-2)\partial^c\Phi_{(S)}\partial_c\ln r 
	+ \frac{2(D-4)\Lambda}{(D-1)(D-2)}\Phi_{(S)}\right]
	+ [k^2-(D-2)K]\Phi_{(S)}. 
\end{equation}
Therefore, the projected Einstein equation $E_{ab}=0$ is equivalent to
the statement that $\Delta_{(S)}$ is a solution of 
%
\begin{equation}
 \nabla_a\nabla_b\Delta_{(S)} + 
	\frac{2\Lambda}{(D-1)(D-2)}\gamma_{ab}\Delta_{(S)}=0. 
	\label{eqn:eq-for-Ds}
\end{equation}

Next, let us show that, by redefinition of $\Phi_{(S)}$, 
$\Delta_{(S)}$ can be set to be zero if $k^2[k^2-(D-2)K]\ne 0$. Even
in the case when $k^2=0$ and $K\ne 0$, $\Delta_{(S)}$ can be set to be
of the form $Cr$, where $C$ is a constant. The proof is as follows. It
is easy to show by using (\ref{eqn:const-curv-dec2}) and
(\ref{eqn:eq-for-Ds}) that 
%
\begin{eqnarray}
 \partial_{\pm}\left(\frac{\Psi_1}{r}\right) & = & 0, 
	\nonumber\\
 \Psi_2 & = & (D-2)Kr, 
	\label{eqn:Psi1-Psi2}
\end{eqnarray}
where $\Psi_1$ and $\Psi_2$ are defined by 
%
\begin{eqnarray}
 \Psi_1 & = & r^2\left[-\nabla^2\Delta_{(S)} 
	+ (D-2)\partial^c\Delta_{(S)}\partial_c\ln r 
	+ \frac{2(D-4)\Lambda}{(D-1)(D-2)}\Delta_{(S)}\right], 
	\nonumber\\
 \Psi_2 & = & r^2\left[-\nabla^2r
	+ (D-2)\partial^cr\partial_c\ln r 
	+ \frac{2(D-4)\Lambda}{(D-1)(D-2)}r\right]
\end{eqnarray}
From the first equation of (\ref{eqn:Psi1-Psi2}), $\Psi_1=\tilde{C}r$,
where $\tilde{C}$ is a constant. Therefore, if $k^2[k^2-(D-2)K]\ne 0$
then  
%
\begin{equation}
 r^2\left[-\nabla^2\Phi'_{(s)} 
	+ (D-2)\partial^c\Phi'_{(s)}\partial_c\ln r 
	+ \frac{2(D-4)\Lambda}{(D-1)(D-2)}\Phi'_{(s)}\right]
	+ [k^2-(D-2)K]\Phi'_{(s)} = 0,
\end{equation}
where
%
\begin{equation}
 \Phi'_{(s)} = \Phi_{(S)} 
	- \frac{\Delta_{(S)}}{k^2-(D-2)K} 
	+ \frac{\tilde{C}r}{k^2[k^2-(D-2)K]}. 
\end{equation}
When $k^2=0$ and $K\ne 0$, we can define $\Phi'_{(s)}$ by
%
\begin{equation}
 \Phi'_{(s)} = \Phi_{(S)} + \frac{\Delta_{(S)}}{(D-2)K}
\end{equation}
so that 
%
\begin{equation}
 r^2\left[-\nabla^2\Phi'_{(s)} 
	+ (D-2)\partial^c\Phi'_{(s)}\partial_c\ln r 
	+ \frac{2(D-4)\Lambda}{(D-1)(D-2)}\Phi'_{(s)}\right]
	-(D-2)K\Phi'_{(s)} = Cr, 
\end{equation}
where $C=\tilde{C}/(D-2)K$. It is evident from
(\ref{eqn:const-curv-dec2}) and (\ref{eqn:eq-for-Ds}) that replacement
of $\Phi_{(S)}$ with $\Phi'_{(s)}$ in Eq.~(\ref{eqn:Phi(s)}) does not
alter $F_{ab}$ nor $F$.

Finally, $F_{ab}$ and $F$ are given by (\ref{eqn:Phi(s)}), where
$\Phi_{(S)}$ is a solution of the master equation
%
\begin{equation}
 \nabla^2\Phi_{(S)} 
	- (D-2)\partial^c\Phi_{(S)}\partial_c\ln r 
	- \frac{2(D-4)\Lambda}{(D-1)(D-2)}\Phi_{(S)}
	-\frac{k^2-(D-2)K}{r^2}\Phi_{(S)} + \frac{\Delta_{(S)}}{r^2}
	= 0. 
\end{equation}
When $k^2[k^2-(D-2)K]\ne 0$, $\Delta_{(S)}=0$. When $k^2=0$ and 
$K\ne 0$, $\Delta_{(S)}=Cr$, where $C$ is an arbitrary constant. 
When $k^2-(D-2)K=0$, $\Delta_{(S)}$ is an arbitrary solution of
(\ref{eqn:eq-for-Ds}).

\subsection{Master equation for vector perturbations}


Now let us consider vector perturbations. First, defining two
functions $\Phi_{(V)\pm}$ by 
%
\begin{equation}
 r^{D-4}F_{\pm} = \pm\partial_{\pm}\Phi_{(V)\pm},
\end{equation}
$E_{(LT)}=0$ is rewritten as
%
\begin{equation}
 \partial_+\partial_-(\Phi_{(V)+}-\Phi_{(V)-}) = 0,
\end{equation}
where $E_{(LT)}$ is given by (\ref{eqn:Ein-eq-form-vec}). Thus, there
are functions $f_5(x_-)$ and $f_6(x_+)$ such that 
%
\begin{equation}
 \Phi_{(V)+}-\Phi_{(V)-} = f_5(x_-) - f_6(x_+). 
\end{equation}
and that we can define a master variable $\Phi_{(V)}$ by 
%
\begin{eqnarray}
 \Phi_{(V)} = \Phi_{(V)+} - f_5(x_-) = \Phi_{(V)-} - f_6(x_+). 
\end{eqnarray}
With this definition, $F_{\pm}$ are expressed as
%
\begin{equation}
 r^{D-4}F_{\pm} = \pm\partial_{\pm}\Phi_{(V)},
\end{equation}
or covariantly,
%
\begin{equation}
 r^{D-4}F_{a} = \epsilon_a^{\ b}\partial_b\Phi_{(V)},
	\label{eqn:Fa-Phiv}
\end{equation}
where $\epsilon_{ab}$ is the Levi-Civita tensor defined by
%
\begin{equation}
 \epsilon_{01} = -\epsilon_{10} = \sqrt{|\det\gamma_{ab}|},\quad 
 \epsilon_{00} = \epsilon_{11} = 0. 
\end{equation}
Note that $\nabla_c\epsilon_{ab}=0$.


Next, by substituting (\ref{eqn:Fa-Phiv}) into
(\ref{eqn:Ein-eq-form-vec}), we obtain
%
\begin{equation}
 \epsilon_b^{\ a}E_{(T)a} = \nabla_b\left\{
	r^D\nabla^a[r^{-(D-2)}\nabla_a\Phi_{(V)}]
	-[k^2-(D-3)K]\Phi_{(V)}\right\}, 
\end{equation}
where we have used the identity 
$\epsilon_b^{\ a}\epsilon_{a'}^{\ c}
=\gamma_{ba'}\gamma^{ac}-\delta_b^c\delta^a_{a'}$. 
Thus, the projected Einstein equation $E_{(T)a}=0$ is equivalent to
the following master equation. 
%
\begin{equation}
 r^{D-2}\nabla^a[r^{-(D-2)}\nabla_a\Phi_{(V)}]
	-\frac{k^2-(D-3)K}{r^2}\Phi_{(V)} 
	+ \frac{\Delta_{(V)}}{r^2} = 0, 
\end{equation}
where $\Delta_{(V)}$ is a constant. Note that, when $k^2\ne (D-3)K$,
$\Delta_{(V)}$ can be set to be zero by redefinition of $\Phi_{(V)}$.


\section{Summary and Discussion}
	\label{sec:summary}


In summary, we have investigated classical perturbations of
$D$-dimensional maximally-symmetric spacetimes (Minkowski, deSitter,
and anti-deSitter spacetimes). We have decomposed the background
spacetime into a family of $n$-dimensional constant-curvature spaces
and have expanded gravitational perturbations by harmonics on the
constant-curvature space. After analyzing gauge transformation, we
constructed gauge-invariant variables. Those can be considered as
scalar fields $F$ and $F_{(T)}$, the vector field $F_a$, and the
symmetric second-rank tensor fields $F_{ab}$ in $(D-n)$-dimensional
spacetime.

When $D-n=2$, we have shown that the tensor field $F_{ab}$ and the
vector field $F_a$ can be described by scalar master
variables. Namely, $F_{ab}$ as well as $F$ are given by
(\ref{eqn:Phi(s)}), and $F_a$ is given by
(\ref{eqn:Fa-Phiv}). Therefore, in this case, we can investigate the 
gauge-invariant perturbations by analyzing the master scalar variables 
$\Phi_{(S)}$ and $\Phi_{(V)}$, and the scalar $F_{(T)}$. These scalar
fields obey the following master equations. 
%
\begin{equation}
 r^{\alpha+\beta}\nabla^a[r^{-\alpha}\nabla_a(r^{-\beta}\Phi)]
	- (k^2+\gamma K)r^{-2}\Phi + \Delta r^{-2} = 0,
\end{equation}
where $\Phi$ represents $\Phi_{(S)}$, $\Phi_{(V)}$ or $F_{(T)}$, and
$K$ is the curvature constant of the $(D-2)$-dimensional
constant-curvature space. The constants ($\alpha$, $\beta$, $\gamma$)
are given by Table~\ref{table:alpha-beta-gamma}, and $\Delta$ is a
constant or a function given by Table~\ref{table:Delta}.


In $4$-dimension ($D=4$), there is a choice such that $\alpha=0$ for
both $\Phi_{(S)}$ and $\Phi_{(V)}$, and there are no degrees of
freedom of $F_{(T)}$ since $T_{(T)ij}\equiv 0$ for $n=2$. (See the
last paragraph of Appendix~\ref{app:harmonics}.) Thus, the result of
this paper is consistent with the master equations given in
Refs.~\cite{KIF,II} for $D=4$, $K=1$.


Here, we mention again that, for $k^2=0$, $F_{ab}$ and $F$
are not gauge invariant variables but gauge dependent variables since 
$V_{(L)i}\equiv 0$ and $T_{(LL)ij}\equiv 0$. Similarly, for a special
value of $k$ such that $T_{(LT)ij}\equiv 0$, $F_{a}$ is not a gauge
invariant variable but a gauge dependent variable. 
The corresponding gauge transformations are 
%
\begin{eqnarray}
 F_{ab} & \to & 
	F_{ab} - \nabla_a\xi_b - \nabla_b\xi_a
	\qquad (k^2=0), \nonumber\\
 F & \to & 
	F_{(Y)} - \gamma^{ab}\xi_a\partial_br^2 
	\qquad (k^2=0),
	\label{eqn:gauge-tr-Fab-F}
\end{eqnarray}
and 
%
\begin{equation}
 F_{a} \to 
	F_{a} - r^2\partial_a(r^{-2}\xi_{(T)})
	\qquad (\mbox{for}\ k\ \mbox{such that}\ T_{(LT)ij}\equiv 0).
	\label{eqn:gauge-tr-Fa}
\end{equation}
The gauge transformation (\ref{eqn:gauge-tr-Fab-F}) for $F_{ab}$ and
$F$ can be considered as the $(D-n)$-dimensional gauge transformation, 
provided that $F_{ab}$ and $F$ are considered as perturbations of
$\gamma_{ab}$ and $r^2$, respectively. 
On the other hand, since $T_{(LT)ij}\equiv 0$ implies that $V_{(T)i}$
is a Killing vector field of the metric $\Omega_{ij}$, from general
arguments of the Kaluza-Klein theory the vector field $F_a$
($=h_{(T)a}$) for such a value of $k$ can be considered as a gauge
field with the gauge group of the isometry of
$\Omega_{ij}$. Correspondingly, (\ref{eqn:gauge-tr-Fa}) can be
considered as the gauge transformation of the gauge field. For all
other values of $k$, off course, $F_{ab}$, $F$, $F_{a}$ and $F_{(T)}$
are gauge invariant variables.


As already explained in Sec.~\ref{sec:intro}, this paper may be
considered as the first step towards the derivation of the
integro-differential equations for cosmological perturbations in the
brane-world scenario. In this respect, the next step, which is now
under investigation, is simplification of Israel's junction 
condition~\cite{Israel} for the master variables given in this paper.


Now let us discuss about a gauge choice which might be convenient in
some cases. From the gauge transformations (\ref{eqn:gauge-tr}), by
choosing $\xi_a$, $\xi_{(T)}$ and $\xi_{(L)}$ as 
%
\begin{eqnarray}
 \xi_{(T)} & = & h_{(LT)}, \nonumber\\
 \xi_{(L)} & = & h_{(LL)}, \nonumber\\
 \xi_a & = & h_{(L)a} - r^2\partial_a(r^{-2}h_{(LL)}),
	\nonumber\\ 
 \xi_{(L)} & = &  -\frac{n}{2k^2}(h_{(Y)}
	-\gamma^{ab}h_{(L)a}\partial_b r^2)	\qquad 
	(\mbox{for}\ k (\ne 0)\ 
	\mbox{such that}\ T_{(LL)ij}\equiv 0).
	\label{eqn:gauge-tr-RW}
\end{eqnarray}
we can always make a gauge transformation such that
%
\begin{eqnarray} 
 h_{(LT)} & \rightarrow & 0, \nonumber\\
 h_{(LL)} & \rightarrow & 0, \nonumber\\
 h_{(L)a} & \rightarrow & 0, \nonumber\\
 h_{(Y)}  & \rightarrow & 0	\qquad 
	(\mbox{for}\ k (\ne 0)\ 
	\mbox{such that}\ T_{(LL)ij}\equiv 0). 
\end{eqnarray}
This gauge choice was adopted in Ref.~\cite{UMM} for $K=1$ in a
different context and may be considered as a generalization of the
so-called Regge-Wheeler gauge~\cite{RW}. 
(For $n=2$, $T_{(T)ij}\equiv 0$ and there is no degrees of freedom of
$h_{(T)}$. See the last paragraph of Appendix~\ref{app:harmonics}.) 
The remaining gauge transformation is
equivalent to the $(D-n)$-dimensional gauge transformation
%
\begin{eqnarray}
 h_{ab} & \to & 
	h_{ab} - \nabla_a\xi_b - \nabla_b\xi_a
	\qquad (k^2=0), \nonumber\\
 h_{(Y)} & \to & 
	h_{(Y)} - \gamma^{ab}\xi_a\partial_br^2 
	\qquad (k^2=0),
\end{eqnarray}
and 
%
\begin{equation}
 h_{(T)a} \to 
	h_{(T)a} - r^2\partial_a(r^{-2}\xi_{(T)})
	\qquad (\mbox{for}\ k\ \mbox{such that}\ T_{(LT)ij}\equiv 0).
	\label{eqn:gauge-tr-hT}
\end{equation}
The vector field $h_{(T)a}$ for $k$ such that $T_{(LT)ij}\equiv 0$ can
be considered as a gauge field with the gauge group of the isometry of
$\Omega_{ij}$, and (\ref{eqn:gauge-tr-hT}) can be considered as the
gauge transformation of the gauge field.


Although the above generalized Regge-Wheeler gauge might be convenient
for some purposes, it seems inconvenient to adopt it when we analyze
perturbations of the brane world. In fact, in general the brane is not
located at $r=R(t)$ in this gauge even if we assume that the
trajectory of the brane is given by $r=R(t)$ in the unperturbed
spacetime. In order to show this, first, let us adopt a Gaussian gauge
in a neighborhood of the world volume of the brane for a moment. Next,
let us perform an infinitesimal gauge transformation so that the
transformed metric perturbation satisfies the generalized
Regge-Wheeler gauge. The infinitesimal gauge transformation is given
by Eq.~(\ref{eqn:gauge-tr-RW}), provided that $h$'s in the right hand 
side are calculated in the Gaussian gauge. Thus, it is easily seen
that 
%
\begin{equation}
 \xi_w =   - r^2\partial_w(r^{-2}h_{(LL)}), 
\end{equation}
where $h_{(LL)}$ in the right hand side is calculated in the Gaussian
gauge, and $w$ is a coordinate corresponding to the geodesic distance
from the brane. Therefore, the displacement of the brane in the
generalized Regge-Wheeler gauge ($\xi_w$ estimated at the brane) is
not zero in general. (cf. Ref.~\cite{GT}) 
In this respect, is is not convenient to adopt the generalized
Regge-Wheeler gauge for the brane world: when one considers spacetime
with singularities such as a domain wall or the brane, perturbative
treatment as well as the variational principle may break down unless
the displacement of the singularities vanishes~\cite{KIF,Hayward}. 
Therefore, if we would prefer to gauge-fixing method rather than the 
gauge-invariant formalism, we have to modify the generalized
Regge-Wheeler gauge slightly so that the displacement of the brane 
vanishes. Otherwise, we have to introduce degrees of freedom for
the displacement of the brane explicitly, and have to consider the
consistent gauge transformation of it so that the gauge transformation 
does not change the physical position of the brane. 
Although the modification of the generalized Regge-Wheeler gauge may
be achieved by allowing non-zero value of $h_{(L)a}$. it seems that in
this modified gauge the analysis becomes complicated. Therefore, the
gauge-invariant formalism developed in this paper seems better than
gauge-fixing method for the analysis of perturbations of the brane
world.


Finally, we suggest a possible generalization of the formalism
developed in this paper. It seems possible to generalize the formalism
to more general background spacetimes. In particular, generalization
to Schwarzschild-AdS spacetime~\cite{Birmingham} is of physical
interests since bulk geometry of cosmological solutions in the
brane-world scenario are Schwarzschild-AdS spacetime in
general~\cite{MSM}.

\begin{acknowledgments}

The author would like to thank Professor W. Israel for continuing
encouragement. He would be grateful to Dr. T. Shiromizu and Professor
M. Sasaki for discussions. This work was supported by the CITA
National Fellowship and the NSERC operating research grant. 

\end{acknowledgments}


\appendix


\section{Relations among three kinds of covariant derivatives}
	\label{app:derivatives}

For an arbitrary (not necessarily symmetric) 2-tensor $T_{MN}$, 
%
\begin{eqnarray}
 T_{ab;cd} & = & \nabla_d\nabla_c T_{ab}, \nonumber\\
 T_{ab;ci} & = & \nabla_c\partial_i T_{ab} 
	- \partial_iT_{ab}\partial_c\ln r 
	- (\nabla_cT_{ib}-2T_{ib}\partial_c\ln r)\partial_a\ln r
	- (\nabla_cT_{ai}-2T_{ai}\partial_c\ln r)\partial_b\ln r, 
	\nonumber\\
 T_{ab;ic} & = & T_{ab;ci} 
	- (T_{ai}\nabla_c\nabla_b\ln r+T_{ib}\nabla_c\nabla_a\ln r), 
 	\nonumber\\
 T_{ai;bc} & = & \nabla_c\nabla_bT_{ai}-\nabla_bT_{ai}\partial_c\ln r
	- (\nabla_cT_{ai}-T_{ai}\partial_c\ln r)\partial_b\ln r
	- T_{ai}\nabla_c\nabla_b\ln r, 
	\nonumber\\
 T_{ab;ij} & = & r^2\Omega_{ij}
	[\nabla_cT_{ab}-T_{ac}\partial_b\ln r 
	- T_{cb}\partial_a\ln r]\partial^c\ln r
	+ D_jD_iT_{ab}	\nonumber\\
 & & 	- (D_iT_{aj}+D_jT_{ai})\partial_b\ln r 
	- (D_iT_{jb}+D_jT_{ib})\partial_a\ln r 
	+ (T_{ij}+T_{ji})\partial_a\ln r\partial_b\ln r, 
	\nonumber\\
 T_{ai;bj} & = & 
	r^2\Omega_{ij}(\nabla_bT_{ac}
	-T_{ac}\partial_b\ln r)\partial^c\ln r
	+\nabla_bD_jT_{ai} - 2 D_jT_{ai}\partial_b\ln r
	- (\partial_bT_{ji}-3T_{ji}\partial_b\ln r)\partial_a\ln r, 
	\nonumber\\
 T_{ai;jb} & = & 
	r^2\Omega_{ij}(\nabla_bT_{ac}\partial^c\ln r
	+ T_{ac}\nabla_b\nabla^c\ln r)
	+\nabla_bD_jT_{ai} - 2 D_jT_{ai}\partial_b\ln r \nonumber\\
 & & 	- (\partial_bT_{ji}-2T_{ji}\partial_b\ln r)\partial_a\ln r
	- T_{ji}\nabla_b\nabla_a\ln r, 
	\nonumber\\
 T_{ij;ab} & = & \nabla_b\nabla_aT_{ij}
	- 2\partial_aT_{ij}\partial_b\ln r
	- 2\partial_bT_{ij}\partial_a\ln r
	+ 2T_{ij}(2\partial_a\ln r\partial_b\ln r
	- \nabla_b\nabla_a\ln r), 
 	\nonumber\\
 T_{ij;ka} & = & 
	r^2\Omega_{ki}[\nabla_aT_{bj}\partial^b\ln r 
	+ T_{bj}(\nabla_a\nabla^b\ln r - 
	\partial_a\ln r\partial^b\ln r)]	\nonumber\\
 & & 	+ r^2\Omega_{jk}[\nabla_aT_{ib}\partial^b\ln r 
	+ T_{ib}(\nabla_a\nabla^b\ln r 
	- \partial_a\ln r\partial^b\ln r)]
 	+ \partial_aD_kT_{ij} - 3D_kT_{ij}\partial_a\ln r, 
	\nonumber\\
 T_{ij;ak} & = & 
	r^2\Omega_{ik}(\nabla_aT_{bj}\partial^b\ln r 
	- 2T_{bj}\partial_a\ln r\partial^b\ln r)
	+ r^2\Omega_{jk}(\nabla_aT_{ib}\partial^b\ln r 
	- 2T_{ib}\partial_a\ln r\partial^b\ln r) \nonumber\\
 & &  	+ \partial_aD_kT_{ij} - 3D_kT_{ij}\partial_a\ln r, 
	\nonumber\\
 T_{ai;jk} & = & r^2(\Omega_{ik}\partial_jT_{ab} + 
	\Omega_{ij}\partial_kT_{ab})\partial^b\ln r
	+ r^2\Omega_{jk}
	(\nabla_bT_{ai}-T_{ai}\partial_b\ln  r)\partial^b\ln r
	- r^2(\Omega_{jk}T_{bi}+\Omega_{ij}T_{kb})
	\partial_a\ln r\partial^b\ln r	\nonumber\\
 & & 	- r^2\Omega_{ik}
	(T_{aj}\partial_b\ln r+T_{jb}\partial_a\ln r)
	\partial^b\ln r
	+ D_kD_jT_{ai}
	-(D_kT_{ji}+D_jT_{ki})\partial_a\ln r,
	\nonumber\\
 T_{ij;kl} & = & r^4(\Omega_{ik}\Omega_{jl}+\Omega_{il}\Omega_{jk})
	T_{ab}\partial^a\ln r\partial^b\ln r
	+ r^2(\Omega_{ik}D_lT_{aj}+\Omega_{jl}D_kT_{ia}
	+\Omega_{jk}D_lT_{ia}+\Omega_{il}D_kT_{aj})\partial^a\ln r
	\nonumber\\
 & & 	+r^2\Omega_{kl}
	(\partial_aT_{ij}-2T_{ij}\partial_a\ln r)\partial^a\ln r
	-r^2(\Omega_{il}T_{kj}+\Omega_{jl}T_{ik})
	\partial_a\ln r\partial^a\ln r 
	+D_lD_kT_{ij}. 
\end{eqnarray}

By using these, we can show the following equations, which are useful
in Sec.~\ref{sec:Einstein-eq}.
%
\begin{eqnarray}
 & & -T_{ab\ \ ;L}^{\ \ ;L} + T_{a\ \ ;Lb}^{\ L}
	+ T_{b\ \ ;La}^{\ L} - T^L_{\ L;ab}\nonumber\\
 & & \quad = 
	-\nabla^2T_{ab}+\nabla_a\nabla^cT_{cb}+\nabla_b\nabla^cT_{ca}
	-\nabla_a\nabla_b(T_{cd}\gamma^{cd})
	+n(\nabla_aT_{bc}+\nabla_bT_{ac}-\nabla_cT_{ab})
	\partial^c\ln r	\nonumber\\
 & & \qquad
	+ n(T_{ac}\partial_b\ln r+T_{cb}\partial_a\ln r)
	\partial^c\ln r
	+ n(T_{ac}\nabla_b\nabla^c\ln r
	+T_{bc}\nabla_a\nabla^c\ln r) - r^{-2}D^2T_{ab}
	+ r^{-2}(\nabla_aD^iT_{bi}+\nabla_bD^iT_{ai})
	\nonumber\\
 & & \qquad
	- r^{-2}\nabla_a\nabla_b(\Omega^{ij}T_{ij})
	+ r^{-2}[\partial_a(\Omega^{ij}T_{ij})\partial_b\ln r
	+\partial_b(\Omega^{ij}T_{ij})\partial_a\ln r]
	-2r^{-2}(\Omega^{ij}T_{ij})\partial_a\ln r\partial_b\ln r
	\nonumber\\
 & &  -T_{ai\ \ ;L}^{\ \ ;L} + T_{a\ \ ;Li}^{\ L}
	+ T_{i\ \ ;La}^{\ L} - T^L_{\ L;ai}\nonumber\\
 & & \quad = \nabla^b\partial_iT_{ab} 
	- \partial_a\partial_i(T_{bc}\gamma^{bc})
	+ (n-2)\partial_iT_{ab}\partial^b\ln r
	+ \partial_i(T_{bc}\gamma^{bc})\partial_a\ln r
	- \nabla^2T_{ai} + \nabla_a\nabla^bT_{ib}
	\nonumber\\
 & & \qquad 
	+\nabla_a[(n-1)T_{ib}+T_{bi}]\partial^b\ln r
	- (n-2)\nabla^bT_{ai}\partial_b\ln r 
	- 2 \nabla^bT_{ib}\partial_a\ln r
	+ T_{ai}(n\partial^b\ln r\partial_b\ln r
	+ \nabla^2\ln r)
	\nonumber\\
 & & \qquad 
	+ [(n-1)T_{ib}+T_{bi}]\nabla_a\nabla^b\ln r 
	- (n-2)(2T_{ib}-T_{bi})\partial_a\ln r\partial^b\ln r
	+ r^{-2}D_iD^jT_{aj} - r^{-2}D^2T_{ai}
	\nonumber\\
 & & \qquad 
	+ r^{-2}\partial_aD^jT_{ij} 
	- r^{-2}\partial_a\partial_i(\Omega^{jk}T_{jk})
	- 2 r^{-2}D^j(2T_{ij}-T_{ji})\partial_a\ln r
	+ 2 r^{-2}\partial_i(\Omega^{jk}T_{jk})\partial_a\ln r
	\nonumber\\
 & &  -T_{ij\ \ ;L}^{\ \ ;L} + T_{i\ \ ;Lj}^{\ L}
	+ T_{j\ \ ;Li}^{\ L} - T^L_{\ L;ij}\nonumber\\
 & & \quad = r^2\Omega_{ij}\left[
	2 \nabla^aT_{ba}\partial^b\ln r 
	- \partial_a(T_{bc}\gamma^{bc})\partial^a\ln r
	+ 2 (n-1)T_{ab}\partial^a\ln r\partial^b\ln r\right]
	- D_iD_j(T_{ab}\gamma^{ab})
	\nonumber\\
 & & \qquad 
	+ \nabla^a(D_iT_{ja}+D_jT_{ia})
	+ \{ D_i[(n-1)T_{ja}-T_{aj}]+D_j[(n-3)T_{ia}+T_{ai}]\}
	\partial^a\ln r
	+2 \Omega_{ij}D^kT_{ak}\partial^a\ln r
	\nonumber\\
 & & \qquad 
	-\nabla^2T_{ij} 
	-(n-4)\partial_aT_{ij}\partial^a\ln r
	-\Omega_{ij}\partial_a(T_{kl}\Omega^{kl})\partial^a\ln r
	+ 2[(n-1)\partial^a\ln r\partial_a\ln r +\nabla^2\ln r]T_{ij}
	\nonumber\\
 & & \qquad
	- r^{-2}D^2T_{ij} 
	+ r^{-2}(D_iD^kT_{jk}+D_jD^kT_{ik})
	- r^{-2}D_jD_i(T_{kl}\Omega^{kl}),
	\nonumber\\
 & &  T^{L\ ;L'}_{\ L\ ;L'} - T^{LL'}_{\quad\ ;LL'}\nonumber\\
 & & \quad = \nabla^2(T_{ab}\gamma^{ab}) 
	- \nabla^b\nabla^aT_{ab}
	+n[\partial_a(T_{bc}\gamma^{bc})-\nabla^b(T_{ab}+T_{ba})]
	\partial^a\ln r
	+ r^{-2}D^2(T_{ab}\gamma^{ab})
	\nonumber\\
 & & \qquad 
	-n T_{ab}(n\partial^a\ln r\partial^b\ln r 
	+ \nabla^a\nabla^b\ln r)
	- r^{-2}\nabla^aD^i(T_{ai}+T_{ia})
	- (n-1)r^{-2}D^i(T_{ai}+T_{ia})\partial^a\ln r
	\nonumber\\
 & & \qquad 
	+ r^{-2}\nabla^2(T_{ij}\Omega^{ij})
	+ (n-3)r^{-2}\partial_a(T_{ij}\Omega^{ij})\partial^a\ln r
	- r^{-2}(T_{ij}\Omega^{ij})
	[(n-2)\partial^a\ln r\partial_a\ln r + \nabla^2\ln r]
	\nonumber\\
 & & \qquad
	+ r^{-4}D^2(T_{ij}\Omega^{ij})
	- r^{-4}D^iD^jT_{ji}. 
	\label{eqn:2ndderivatives}
\end{eqnarray}


\section{Harmonics on constant-curvature space}
	\label{app:harmonics}

In this Appendix we give definitions and basic properties of scalar,
vector and tensor harmonics on a $n$-dimensional constant-curvature
space. Throughout this Appendix we will use the notation that
$\Omega_{ij}$ is the metric of the constant-curvature space and that 
$D_i$ is the covariant derivative compatible with $\Omega_{ij}$. The
curvature tensor of the space is given by Eq.~(\ref{eqn:R-Omega}).

\subsection{scalar harmonics}

The scalar harmonics is supposed to satisfy the following relations.
%
\begin{eqnarray}
 D^2 Y + k^2Y & = & 0,	\nonumber\\
 \int d^nx\sqrt{\Omega}Y Y & = & \delta. 
\end{eqnarray}
Hereafter, $k^2$ denotes continuous eigenvalues for $K=0,-1$~\cite{VS} 
or discrete eigenvalues $k_l^2=l(l+n-1)$ ($l=0,1,\cdots$) for
$K=1$~\cite{RR}, and we omit them in most cases. In this respect, the 
delta $\delta$ in equations above and below represents Dirac's delta
function $\delta^n(k-k')$ for continuous eigenvalues and Kronecker's
delta $\delta_{ll'}\delta_{mm'}$ for discrete eigenvalues, where $m$
(and $m'$) denotes a set of integers. Correspondingly, in the
following arguments, a summation with respect to $k$ should be
understood as integration for $K=0,-1$.

\subsection{vector harmonics}

First, in general, a vector field $v_i$ can be decomposed as 
%
\begin{equation} 
 v_i=v_{(T)i}+\partial_i f , 
\end{equation}
where $f$ is a function and $v_{(T)}$ is a transverse vector field:
%
\begin{equation}
 D^iv_{(T)i}=0 .
\end{equation}

Thus, the vector field $v_i$ can be expanded by using the scalar
harmonics $Y$ and transverse vector harmonics $V_{(T)i}$ as 
%
\begin{equation}
 V_i = \sum_k\left[c_{(T)}V_{(T)i}+c_{(L)}\partial_i Y\right],
	\label{eqn:dY+V}
\end{equation}
where $c_{(T)}$ and $c_{(L)}$ are constants depending on $k$, and the
transverse vector harmonics $V_{(T)i}$ is supposed to satisfy the
following relations.  
%
\begin{eqnarray}
 D^2V_{(T)i} + k^2 V_{(T)i} & = & 0,	\nonumber\\
 D^iV_{(T)i} & = & 0,\nonumber\\
 \int d^nx\sqrt{\Omega}\Omega^{ij}V_{(T)i}V_{(T)j}
	& = & \delta,
\end{eqnarray}
where $k^2$ denotes continuous eigenvalues for $K=0,-1$ or discrete
eigenvalues $k_l^2=l(l+n-1)-1$ ($l=1,2,\cdots$) for $K=1$~\cite{RR},
and we omit them in most cases. From Eq.~(\ref{eqn:dY+V}), it is
convenient to define longitudinal vector harmonics $V_{(L)i}$ by  
%
\begin{equation}
 V_{(L)i} \equiv \partial_i Y. 
\end{equation}

It is easily shown that the longitudinal vector harmonics has the 
following properties. 
%
\begin{eqnarray}
 D^2V_{(L)i} + [k^2-(n-1)K]V_{(L)i} & = & 0,\nonumber\\ 
 D^iV_{(L)i} & = & -k^2Y,\nonumber\\
 D_{[i}(V_{(L)j]} & = & 0,\nonumber \\
 \int d^nx\sqrt{\Omega}\Omega^{ij}V_{(L)i}V_{(L)j}
	& = & k^2\delta,\nonumber\\
 \int d^nx\sqrt{\Omega}\Omega^{ij}V_{(T)i}V_{(L)j} 
	& = & 0.  
\end{eqnarray}

\subsection{Tensor harmonics}

First, in general, a symmetric second-rank tensor field $t_{ij}$ can
be decomposed as
%
\begin{equation}
 t_{ij}=t_{(T)ij} + D_iv_j+D_jv_i + f\Omega_{ij},
\end{equation}
where $f$ is a function, $v_i$ is a vector field and $t_{(T)ij}$ is
a transverse traceless symmetric tensor field:
%
\begin{eqnarray}
 t_{(T)i}^i & = & 0,\nonumber\\
 D^i t_{(T)ij} & = &0. 
	\label{eqn:trasverse-traceless}
\end{eqnarray}

Thus, the tensor field $t_{ij}$ can be expanded by using the vector
harmonics $V_{(T)}$ and $V_{(L)}$, and transverse traceless tensor 
harmonics $T_{(T)}$ as 
%
\begin{equation}
 t_{ij} = \sum_{k}\left[
	c_{(T)}T_{(T)ij}+c_{(LT)}(D_iV_{(T)j}+D_jV_{(T)i})
	+ c_{(LL)}(D_iV_{(L)j}+D_jV_{(L)i})
	+ c_{(Y)}Y\Omega_{ij}\right],
	\label{eqn:dV+T}
\end{equation}
where $c_{(T)}$, $c_{(LT)}$, $c_{(LL)}$ and $c_{(Y)}$ are constants
depending on $k$, and the transverse tensor harmonics $T_{(T)}$ is
supposed to satisfy the following relations. 
%
\begin{eqnarray}
 D^2 T_{(T)ij}+k^2T_{(T)ij} & = & 0,\nonumber\\ 
 T_{(T)i}^i & = & 0,\nonumber\\
 D^iT_{(T)ij} & = & 0,\nonumber\\
 \int d^nx\sqrt{\Omega}\Omega^{ik}\Omega^{jl}T_{(T)ij}T_{(T)kl} 
	& = & \delta, 
\end{eqnarray}
where $k^2$ denotes continuous eigenvalues for $K=0,-1$ or discrete
eigenvalues $k_l^2=l(l+n-1)-2$ ($l=2,3,\cdots$) for $K=1$~\cite{RR},
and we omit them in most cases. From Eq.~(\ref{eqn:dV+T}), it is
convenient to define tensor harmonics $T_{(LT)}$, $T_{(LL)}$, and
$T_{(Y)}$ by 
%
\begin{eqnarray}
 T_{(LT)ij} & \equiv & D_iV_{(T)j}+D_jV_{(T)i}, \nonumber\\
 T_{(LL)ij} & \equiv & D_iV_{(L)j}+D_jV_{(L)i}
	-\frac{2}{n}\Omega_{ij}D^kV_{(L)k}	\nonumber\\
 & = & 2D_iD_jY+\frac{2}{n}k^2\Omega_{ij}Y,	\nonumber\\
 T_{(Y)ij} & \equiv & \Omega_{ij}Y. 
\end{eqnarray}

It is easily shown that these tensor harmonics satisfy the following
properties.
%
\begin{eqnarray}
 D^2 T_{(LT)ij} + [k^2-(n+1)K]T_{(LT)ij} & = & 0,
	\nonumber\\ 
 D^iT_{(LT)ij} & = & -[k^2-(n-1)K]V_{(T)j}, 
	\nonumber\\
 T_{(LT)i}^i & = & 0,
\end{eqnarray}
%
\begin{eqnarray}
 D^2 T_{(LL)ij} + [k^2-2nK]T_{(LL)ij} & = &0 
	,\nonumber\\
 D^iT_{(LL)ij} & = & -\frac{2(n-1)}{n}(k^2-nK)V_{(L)j},
	\nonumber\\
 T_{(LL)i}^i & = & 0,
\end{eqnarray}
and 
%
\begin{eqnarray}
 D^2 T_{(Y)ij} + k^2 T_{(Y)ij} & = & 0,\nonumber\\
 D^i T_{(Y)ij} & = & V_{(L)j},\nonumber\\ 
 T_{(Y)i}^i & = & nY. 
\end{eqnarray}

It is also easy to show the following formulas of integral as well as
the orthogonality between any different types of tensor harmonics.  
%
\begin{eqnarray}
 \int d^nx\sqrt{\Omega}\Omega^{ik}\Omega^{jl}
	T_{(LT)ij}T_{(LT)kl}
 & = & 2[k^2-(n-1)K]\delta, \nonumber\\
 \int d^nx\sqrt{\Omega}\Omega^{ik}\Omega^{jl}
	T_{(LL)ij}T_{(LL)kl}
 & = & \frac{4(n-1)}{n}(k^2-nK)k^2\delta, \nonumber\\
 \int d^nx\sqrt{\Omega}\Omega^{ik}\Omega^{jl}
	T_{(Y)ij}T_{(Y)kl}
 & = & n\delta.
\end{eqnarray}

Finally, we prove that $T_{(T)ij}\equiv 0$ for $n=2$. First, without
loss of generality, we can assume that the metric is of the form
%
\begin{equation}
 \Omega_{ij}dx^idx^j = 2e^{\psi}dzd\bar{z},
\end{equation}
where $\psi$ is a function of a complex coordinate $z$ and its complex 
conjugate $\bar{z}$. In this coordinate system, the
transverse-traceless condition (\ref{eqn:trasverse-traceless}) becomes 
%
\begin{eqnarray}
 t_{(T)z\bar{z}} & = & 0,	\nonumber\\
 \partial_{\bar{z}}t_{(T)zz} & = & 
	\partial_zt_{(T)\bar{z}\bar{z}} = 0. 
\end{eqnarray}
The second equation can be solved to give $t_{(T)zz}=t(z)$ and
$t_{(T)\bar{z}\bar{z}}=\tilde{t}(\bar{z})$, where $t$ and $\bar{t}$
are arbitrary holomorphic and anti-holomorphic functions. Thus,
$t_{ij}$ can be written as follows. 
%
\begin{eqnarray}
 t_{(T)ij} = D_iv_j + D_jv_i + f\Omega_{ij},
\end{eqnarray}
where the vector $v_i$ and the scalar $f$ are defined by 
$v_z\equiv e^{-\psi}\int dze^{\psi}t(z)/2$, 
$v_{\bar{z}}\equiv e^{-\psi}\int d\bar{z}e^{\psi}\tilde{t}(\bar{z})/2$
and $f=-e^{-\psi}(\partial_z v_{\bar{z}}+\partial_{\bar{z}}v_z)$. This
means that any transverse-traceless tensor can be written in terms of
$T_{(LT)ij}$, $T_{(LL)ij}$ and $T_{(Y)ij}$. This completes the proof
that $T_{(T)ij}\equiv 0$ for $n=2$.


\section{General solution of the consistency condition}
	\label{app:solution}

In this appendix, we seek a general solution of (\ref{eqn:ddD=LeD})
for $\tilde{\Lambda}\ne 0$.

First, let us define a new function $X$ by 
%
\begin{equation}
 \Delta = e^{-\phi}\partial_+(e^{\phi}X). 
\end{equation}
Thence, the equation (\ref{eqn:ddD=LeD}) is equivalent to 
%
\begin{equation}
 \partial_+[e^{-\phi}\partial_+(e^{\phi}\partial_-X)] = 0. 
\end{equation}
This can be easily integrated to give
%
\begin{eqnarray}
 \partial_+(e^{\phi}\partial_-X) & = & f_7(x_-)e^{\phi}
	\nonumber\\
 & = & \frac{1}{\tilde{\Lambda}}f_7(x_-)\partial_+\partial_-\phi,
\end{eqnarray}
where $f_7(x_-)$ is an arbitrary function and we have used the last
equation of (\ref{eqn:const-curv-dec2}) to obtain the last line. This
equation can also be integrated to give
%
\begin{eqnarray}
 \partial_-X & = & e^{-\phi}\left[
	\frac{1}{\tilde{\Lambda}}f_7(x_-)\partial_-\phi + f_8(x_-)
	\right]	\nonumber\\
 & = & -\frac{1}{\tilde{\Lambda}}\partial_-[e^{-\phi}f_7(x_-)]
	+ e^{-\phi}\left[
	\frac{1}{\tilde{\Lambda}}\partial_-f_7(x_-)+ f_8(x_-)\right],
\end{eqnarray}
where $f_8(x_-)$ is also an arbitrary function. Hence,
%
\begin{equation}
 X = -\frac{1}{\tilde{\Lambda}}e^{-\phi}f_7(x_-)
	+\int dx_-  e^{-\phi}\left[
	\frac{1}{\tilde{\Lambda}}\partial_-f_7(x_-)+ f_8(x_-)\right]
	+ f_9(x_+),
\end{equation}
where $f_9(x_+)$ is an arbitrary function. Therefore the general
solution of (\ref{eqn:ddD=LeD}) can be written as
%
\begin{equation}
 \Delta = \Delta_+ + \Delta_-,
\end{equation}
where
%
\begin{eqnarray}
 \Delta_+ & = & e^{-\phi}\partial_+[e^{\phi}C^+(x_+)], 
	\nonumber\\
 \Delta_- & = & e^{-\phi}\partial_+\left[
	e^{\phi}\int dx_-e^{-\phi}C(x_-)\right], 
\end{eqnarray}
where $C^+$ and $C$ are arbitrary functions.

Next, let us show that $\Delta_-$ can be rewritten as
$e^{-\phi}\partial_-[e^{\phi}C^-(x_-)]$ by some function
$C^-$. This is easily done in a particular coordinate system 
as we shall show below. Hence, let us see that the form of $\Delta_-$
and that of $e^{-\phi}\partial_-[e^{\phi}C^-(x_-)]$ are invariant
under a coordinate transformation~\footnote{
Thanks to the $2$-dimensional version of the uniqueness of the
constant-curvature spacetime~\cite{Weinberg}
(cf. the last equation of (\ref{eqn:const-curv-dec})), the metric
$\gamma_{ab}$ for different value of $K$ can also be obtained by a
coordinate transformation from the metric $\gamma_{ab}$ in a particular
coordinate system for a particular value of $K$, provided that
$\Lambda$ is common. However, the explicit expression of $r$ will
change if $\gamma_{ab}$ is expressed in the common form.}. 
In fact, under a general coordinate transformation
$x_{\pm}\to\tilde{x}_{\pm}=\tilde{x}_{\pm}(x_{\pm})$ between 
double-null coordinate systems, these forms are invariant: 
%
\begin{eqnarray}
 e^{-\phi}\partial_+\left[ 
	e^{\phi}\int dx_-e^{-\phi}C(x_-)\right] & = & 
 e^{-\tilde{\phi}}\tilde{\partial}_+\left[ 
	e^{\tilde{\phi}}\int d\tilde{x}_-
	e^{-\tilde{\phi}}\tilde{C}(\tilde{x}_-)\right], \nonumber\\
 e^{-\phi}\partial_-[e^{\phi}C^-(x_-)] & = & 
	e^{-\tilde{\phi}}\tilde{\partial}_-
	[e^{\tilde{\phi}}\tilde{C}^-(\tilde{x}_-)],
\end{eqnarray}
where
$e^{\tilde{\phi}}=e^{\phi}(dx_+/d\tilde{x}_+)(dx_-/d\tilde{x}_-)$, 
$\tilde{\partial}_{\pm}
=(\partial /\partial\tilde{x}_{\pm})_{\tilde{x}_{\mp}}$ and  
%
\begin{eqnarray}
 \tilde{C}(\tilde{x}_-) & = & C(x_-)
	\left(\frac{dx_-}{d\tilde{x}_-}\right)^2,\nonumber\\
 \tilde{C}^-(\tilde{x}_-) & = & C^-(x_-)
	\frac{d\tilde{x}_-}{dx_-}.
\end{eqnarray}

Now let us show in a particular coordinate system that $\Delta_-$ can
actually be rewritten as $e^{-\phi}\partial_-[e^{\phi}C^-(x_-)]$ by
some function $C^-$. For this purpose, it seems the easiest to
consider a coordinate system in which
%
\begin{equation}
 e^{\phi} = -\frac{(D-1)(D-2)}{\Lambda (x_+-x_-)^2}. 
\end{equation}
In this coordinate, it is easy to show by integrations by part that 
%
\begin{equation}
 \Delta_- = e^{-\phi}\partial_-[e^{\phi}C^-(x_-)], 
\end{equation}
where $C^-(x_-)$ is defined by
%
\begin{equation}
 \partial_-^3 C^-(x_-) = \frac{2\Lambda}{(D-1)(D-2)}C(x_-).
\end{equation}

Finally, we have shown that a general solution of (\ref{eqn:ddD=LeD})
for $\tilde{\Lambda}\ne 0$ is 
%
\begin{equation}
 \Delta = e^{-\phi}\partial_+[e^{\phi}C^+(x_+)]
	+ e^{-\phi}\partial_-[e^{\phi}C^-(x_-)], 
\end{equation}
where $C^{\pm}$ are arbitrary functions.




%
\begin{table}
\caption{Two sets of values of ($\alpha$, $\beta$, $\gamma$)}
	\label{table:alpha-beta-gamma}
\begin{center}
\begin{tabular}{|c||c|c|c||c|c|c|} 
 $\Phi$ & $\alpha$ & $\beta$ & $\gamma$ 
	& $\alpha$ & $\beta$ & $\gamma$ \\ \hline
 $\Phi_{(S)}$ & $D-4$ & $1$ & $0$ 
	& $-(D-6)$ & $D-4$ & $2(D-5)$ \\
 $\Phi_{(V)}$ & $D-2$ & $0$ & $-(D-3)$ 
	& $-(D-4)$ & $D-3$ & $D-3$ \\
 $F_{(T)}$ & $D$ & $-(D-3)$ & $-2(D-2)$ 
	& $-(D-2)$ & $2$ & $2$ \\ 
\end{tabular}
\end{center}
\end{table}

%
\begin{table}
\caption{$\Delta$}
	\label{table:Delta}
\begin{center}
\begin{tabular}{|c|c||c|}
 $\Phi$ & $k$ & $\Delta$  \\ \hline
 $\Phi_{(S)}$ & $k^2[k^2-(D-2)K]\ne 0$ & $0$ \\
 	& $k^2=0$ and $K\ne 0$ & Constant $\times\ r$ \\
 	& $k^2=(D-2)K$ 
	& Solution $\Delta_{(S)}$ of (\ref{eqn:eq-for-Ds}) \\
 $\Phi_{(V)}$ & $k^2\ne (D-3)K$ & $0$ \\
 	& $k^2=(D-3)K$ & Constant \\
 $F_{(T)}$ & ${}^{\forall}k$ & $0$ \\
\end{tabular}
\end{center}
\end{table}


\begin{references}
\bibitem{Maldacena}
J.~Maldacena, Adv. Theor. Math. Phys., {\bf 2}, 231(1998).
\bibitem{AdS-CFT}
O.~Aharony, S.~S.~Gubser, J.~Maldacena, H.~Ooguri and Y.~Oz,
hep-th/9905111 and reference therein. 
\bibitem{RS}
L.~Randall and R.~Sundrum, Phys.Rev.Lett. {\bf 83}, 4690 (1999)
[hep-th/9906064]. 
\bibitem{SMS}
T.~Shiromizu, K.~Maeda and M.~Sasaki, gr-qc/9910076.
\bibitem{CG}
A.~Chamblin and G.~W.~Gibbons, hep-th/9909130.
\bibitem{GT}
J.~Garriga and T.~Tanaka, hep-th/9911055.
\bibitem{Tanaka-Montes}
T.~Tanaka and X.~Montes, hep-th/0001092. 
\bibitem{SSM}
M.~Sasaki, T.~Shiromizu and K.~Maeda, hep-th/9912233.
\bibitem{GKR}
S.~B.~Giddings, E.~Katz, and L.~Randall, hep-th/0002091. 
\bibitem{CHR}
A.~Chamblin, S.~W.~Hawking, and H.~ S.~Reall, hep-th/9909205.
\bibitem{EHM}
R.~Emparan, G.~T.~Horowitz and R.~C.~Myers, hep-th/9911043; 
hep-th/9912135.
\bibitem{Nihei}
T.~Nihei, Phys. Lett. {\bf B465} 81 (1999) [hep-ph/9905487].
\bibitem{Kaloper}
N.~Kaloper, Phys. Rev. {\bf D60} 123506 (1999) [hep-th/9905210]. 
\bibitem{KK}
H.~B.~Kim  and H.~D.~Kim, hep-th/9909053. 
\bibitem{Gubser}
S.~S.~Gubser, hep-th/9912001.
\bibitem{CGS}
J.~M.~Cline, C.~Grojean and G.~Servant, Phys. Rev. Lett. 
{\bf 83} 4245 (1999) [hep-ph/9906523].
\bibitem{FTW}
E.~E.~Flanagan, S.~H.~H.~Tye, I.~Wasserman, hep-ph/9910498.
\bibitem{BDEL}
P.~Bin\'{e}truy, C.~Deffayet, U.~Ellwanger and D.~Langlois,
hep-th/9910219. 
\bibitem{Mukohyama}
S.~Mukohyama, Phys. Lett. {\bf B473}, 241 (2000) [hep-th/9911165]. 
\bibitem{Vollic}
D.~N.~Vollick, hep-th/9911181.
\bibitem{Kraus}
P.~Kraus, JHEP {\bf 9912}, 011 (1999) [hep-th/9910149].
\bibitem{Ida}
D.~Ida, gr-qc/9912002.
\bibitem{MSM}
S.~Mukohyama, T.~Shiromizu and K.~Maeda, hep-th/9912287, to appear in
Phys. Rev. D. 
\bibitem{Birmingham}
D.~Birmingham, Class. Quant. Grav. {\bf 16}, 1197(1999).
\bibitem{Hawking}
S.~W.~Hawking, Commun. Math. Phys. {\bf 43}, 199 (1975).
\bibitem{conservation}
S.~Wands, et. al., astro-ph/0003278. 
\bibitem{GS}
J.~Garriga and M.~Sasaki, hep-th/9912118.
\bibitem{Koyama-Soda}
K.~Koyama and J.~Soda, gr-qc/0001033.
\bibitem{HHR}
S.~W.~Hawking, T.~Hertog, and H.~S.~Reall, hep-th/0003052. 
\bibitem{Weinberg}
For the proof of the uniqueness, see for example, S.~Weinberg, {\it 
Gravitation and cosmology} (John Willey \& Sons, Inc., 1972). 
\bibitem{Hawking-Ellis}
For properties of these spacetimes, see for example, S.~W.~Hawking and
G.~F.~R.~Ellis, {\it The large scale structure of space-time},
(Cambridge University Press, 1973). 
\bibitem{Kodama-Sasaki}
H.~Kodama and M.~Sasaki, Prog. Theor. Phys. Suppl. {\bf 78}, 1
(1984). 
\bibitem{KIF}
H.~Kodama, H.~Ishihara and Y.~Fujiwara, Phys. Rev. {\bf D50}, 7292
(1994). 
\bibitem{II}
A.~Ishibashi and H.~Ishihara, Phys. Rev. {\bf D56}, 3446 (1997). 
\bibitem{Israel}
W.~Isarel, Nuovo Cim. {\bf 44B}, 1 (1966). 
\bibitem{UMM}
K.~Uzawa, Y.~Morisawa and S.~Mukohyama, gr-qc/9912108. 
\bibitem{RW}
T.~Regge and J.~A.~Wheeler, Phys. Rev. {\bf 108}, 1063 (1957). 
\bibitem{Hayward}
G.~Hayward and J.~Louko, Phys. Rev. {\bf D42}, 4032 (1990). 
\bibitem{VS}
N.~Y.~Vilenkin and Y.~A.~Smorodinski\v{i}, Sov. Phys. JETP, {\bf 19},
1209 (1964). 
\bibitem{RR}
M.~A.~Rubin and C.~R.~Ord\'{o}\~{n}ez, J. Math. Phys. {\bf 25}, 2888 
(1984); {\bf 26}, 65 (1985).
\end{references}
\end{document}